%% file: neurips_2021.tex
\documentclass{article}

\PassOptionsToPackage{numbers, compress}{natbib}

    \usepackage[final]{neurips_2021}

\usepackage[utf8]{inputenc} 
\usepackage[T1]{fontenc}    
\usepackage{hyperref}       
\usepackage{url}            
\usepackage{booktabs}       
\usepackage{amsfonts}       
\usepackage{nicefrac}       
\usepackage{microtype}      
\usepackage{xcolor}         
\usepackage{graphicx}
\newcommand*\samethanks[1][\value{footnote}]{\footnotemark[#1]}

\title{Effect of Radiology Report Labeler Quality on Deep Learning Models for Chest X-Ray Interpretation}

\author{%
  Saahil Jain\thanks{Equal Contribution} \\
  Stanford University\\
  \texttt{saahil.jain@cs.stanford.edu} \\
   \And
   Akshay Smit\samethanks[1] \\
   Stanford University \\
   \texttt{akshaysm@cs.stanford.edu} \\
  \And
  Andrew Y. Ng \\
  Stanford University \\
  \texttt{ang@cs.stanford.edu} \\
   \And
   Pranav Rajpurkar \\
   Harvard University \\
   \texttt{pranav\_rajpurkar@hms.harvard.edu} \\
}

\begin{document}

\maketitle

\begin{abstract}
Although deep learning models for chest X-ray interpretation are commonly trained on labels generated by automatic radiology report labelers, the impact of improvements in report labeling on the performance of chest X-ray classification models has not been systematically investigated. We first compare the CheXpert, CheXbert, and VisualCheXbert labelers on the task of extracting accurate chest X-ray image labels from radiology reports, reporting that the VisualCheXbert labeler outperforms the CheXpert and CheXbert labelers. Next, after training image classification models using labels generated from the different radiology report labelers on one of the largest datasets of chest X-rays, we show that an image classification model trained on labels from the VisualCheXbert labeler outperforms image classification models trained on labels from the CheXpert and CheXbert labelers. Our work suggests that recent improvements in radiology report labeling can translate to the development of higher performing chest X-ray classification models.

\end{abstract}

\section{Introduction}
Given the high cost of obtaining expert labels for radiology images, deep learning models for radiology image interpretation are often trained on labels automatically extracted from reports accompanying images. Although automatic radiology report labelers are routinely used to label large chest X-ray datasets \cite{irvin2019chexpert, johnson2019mimiccxrjpg}, the impact of recent improvements in report labeling on the training of automated chest X-ray classification models is unexplored. To determine the utility of such improvements for the development of radiology image interpretation models, we explore how radiology report labeler quality affects the performance of chest X-ray classification models. First, we compare the quality of various radiology report labelers. Second, we show that the reported improvements in radiology report labeling translate to improvements in chest X-ray image interpretation.

\section{Experiments}

\subsection{Data and limitations}
\label{data_section}
We use the CheXpert dataset \cite{irvin2019chexpert}, which contains a training set of 224,316 chest radiographs, a validation set of 200 chest X-ray studies, and a test set of 500 chest X-ray studies. Validation and test set labels are obtained by radiologists annotating X-rays. For our experiments, we evaluate on conditions with at least 10\% prevalence in the test set. We limit our experiments to the CheXpert dataset with three released radiology report labelers and the imaging model from CheXpert \cite{irvin2019chexpert}.

\subsection{Comparison of radiology report labelers}
\label{rad_labeler_comparisons}
We compare the CheXpert \cite{irvin2019chexpert}, CheXbert \cite{smit2020chexbert}, and VisualCheXbert \cite{Jain_2021} labelers on the task of extracting chest X-ray image labels from radiology reports, described in Jain et al. \cite{Jain_2021}. To handle the uncertain classes outputted by the CheXpert and CheXbert labelers, we use two common uncertainty handling strategies in Irvin et al. \cite{irvin2019chexpert}; we either map all the uncertain classes to the negative class (U-Zeros) or to the positive class (U-Ones). We evaluate the performance of each labeler on the CheXpert test set on the different conditions. We also report the weighted average across conditions, where scores are weighted by their prevalence in the test set. We report F1 scores with 95\% confidence intervals generated using the non-parametric percentile bootstrap method with 1000 bootstrap replicates \cite{efron1986bootstrap}.

The VisualCheXbert labeler achieves a statistically significant improvement in average and weighted average F1 scores across conditions compared to the CheXbert and CheXpert labeler approaches, which perform similarly. As shown in Table \ref{labeler_performance}, the VisualCheXbert labeler obtains a weighted average F1 of 0.73 (95\% CI 0.71, 0.75), higher than the weighted average F1 of 0.55 (95\% CI 0.53, 0.58) for the next best approach.


\input{tables/labeler_performance}

\subsection{Comparison of chest X-ray classification models}
\label{cnn_model_comparisons}
We train image classification models using labels generated by different radiology report labelers on the CheXpert training set. For each model (DenseNet-121 architecture), we use 3 TITAN-XP GPUs following the training procedure specified by Irvin et al. \cite{irvin2019chexpert}. We use the best checkpoint, as measured on the validation set, for evaluation. We evaluate the performance of each model on the CheXpert test set, reporting AUROC scores with 95\% confidence intervals generated as described in Section \ref{rad_labeler_comparisons}.


The image classification model trained on labels from the VisualCheXbert labeler achieves a statistically significant improvement in average and weighted average AUROC scores across conditions compared to models trained on labels from the CheXbert and CheXpert labelers, which perform similarly. As shown in Table \ref{cnn_performance}, the model trained on VisualCheXbert labels obtains a weighted average AUROC of 0.87 (95\% CI 0.86, 0.89), higher than the weighted average AUROC of 0.83 (95\% CI 0.81, 0.84) for the next best approach.

\input{tables/cnn_performance}

Given the statistically significant overall improvement achieved by the model trained using the VisualCheXbert labeler, we show that recent improvements in radiology report labeling (Section \ref{rad_labeler_comparisons}) can indeed translate to improvements in chest X-ray classification models (Section \ref{cnn_model_comparisons}).


\medskip

{
\small

\bibliographystyle{unsrt}
\bibliography{references}

}

\end{document}

%% file: tables/labeler_performance.tex
\begin{table}[hbt!]
\centering
\caption{F1 scores with 95\% confidence intervals obtained by report labelers on the CheXpert test set.}
\label{labeler_performance}
\resizebox{\textwidth}{!}{%
\begin{tabular}{llllll}
Condition (n = \#positive) &
  \multicolumn{1}{c}{\begin{tabular}[c]{@{}c@{}}CheXpert \\ U-Zeros\end{tabular}} &
  \multicolumn{1}{c}{\begin{tabular}[c]{@{}c@{}}CheXpert \\ U-Ones\end{tabular}} &
  \multicolumn{1}{c}{\begin{tabular}[c]{@{}c@{}}CheXbert \\ U-Zeros\end{tabular}} &
  \multicolumn{1}{c}{\begin{tabular}[c]{@{}c@{}}CheXbert \\ U-Ones\end{tabular}} &
  \multicolumn{1}{c}{VisualCheXbert} \\ \hline
Atelectasis (n=153)        & 0.30 (0.22, 0.38) & 0.51 (0.44, 0.58) & 0.30 (0.22, 0.38) & 0.51 (0.44, 0.57) & \textbf{0.63 (0.58, 0.69)} \\
Cardiomegaly (n=151)       & 0.44 (0.36, 0.52) & 0.46 (0.38, 0.54) & 0.43 (0.35, 0.51) & 0.46 (0.38, 0.54) & \textbf{0.62 (0.55, 0.67)} \\
Edema (n=78)               & 0.46 (0.36, 0.55) & 0.53 (0.45, 0.61) & 0.49 (0.39, 0.58) & 0.53 (0.44, 0.62) & 0.54 (0.46, 0.62)          \\
Pleural Effusion (n=104)   & 0.65 (0.58, 0.71) & 0.65 (0.59, 0.72) & 0.66 (0.59, 0.72) & 0.66 (0.59, 0.72) & 0.65 (0.58, 0.71)          \\
Enlarged Cardiom. (n=253) & 0.10 (0.05, 0.15) & 0.20 (0.14, 0.26) & 0.12 (0.07, 0.17) & 0.23 (0.17, 0.28) & \textbf{0.73 (0.68, 0.77)} \\
Lung Opacity (n=264)       & 0.69 (0.63, 0.74) & 0.68 (0.63, 0.73) & 0.67 (0.62, 0.72) & 0.67 (0.62, 0.72) & \textbf{0.83 (0.79, 0.86)} \\
Support Devices (n=261)    & 0.85 (0.81, 0.88) & 0.84 (0.81, 0.87) & 0.84 (0.80, 0.88) & 0.84 (0.79, 0.87) & 0.87 (0.84, 0.90)          \\
No Finding (n=62)          & 0.39 (0.28, 0.49) & 0.38 (0.28, 0.48) & 0.36 (0.26, 0.46) & 0.36 (0.27, 0.45) & 0.54 (0.45, 0.62)          \\ \hline
Average                    & 0.48 (0.46, 0.51) & 0.53 (0.51, 0.56) & 0.48 (0.46, 0.51) & 0.53 (0.51, 0.56) & \textbf{0.68 (0.65, 0.70)} \\
Weighted Average           & 0.50 (0.48, 0.53) & 0.55 (0.53, 0.58) & 0.50 (0.48, 0.53) & 0.55 (0.53, 0.58) & \textbf{0.73 (0.71, 0.75)}
\end{tabular}%
}
\end{table}

%% file: tables/cnn_performance.tex
\begin{table}[hbt!]
\centering
\caption{AUROC scores with 95\% confidence intervals obtained by vision models on the CheXpert test set. Column names correspond to respective labelers used.}
\label{cnn_performance}
\resizebox{\textwidth}{!}{%
\begin{tabular}{llllll}
Condition (n = \#positive) &
  \multicolumn{1}{c}{\begin{tabular}[c]{@{}c@{}}CheXpert \\ U-Zeros\end{tabular}} &
  \multicolumn{1}{c}{\begin{tabular}[c]{@{}c@{}}CheXpert \\ U-Ones\end{tabular}} &
  \multicolumn{1}{c}{\begin{tabular}[c]{@{}c@{}}CheXbert \\ U-Zeros\end{tabular}} &
  \multicolumn{1}{c}{\begin{tabular}[c]{@{}c@{}}CheXbert \\ U-Ones\end{tabular}} &
  \multicolumn{1}{c}{VisualCheXbert} \\ \hline
Atelectasis (n=153)        & 0.70 (0.65, 0.75) & 0.75 (0.71, 0.79) & 0.76 (0.71, 0.80) & 0.77 (0.72, 0.81) & \textbf{0.84 (0.81, 0.88)} \\
Cardiomegaly (n=151)       & 0.82 (0.78, 0.85) & 0.79 (0.74, 0.84) & 0.79 (0.75, 0.83) & 0.74 (0.69, 0.79) & 0.83 (0.79, 0.86) \\
Edema (n=78)               & 0.88 (0.85, 0.92) & 0.91 (0.88, 0.94) & 0.88 (0.84, 0.91) & 0.89 (0.85, 0.92) & 0.89 (0.86, 0.92)          \\
Pleural Effusion (n=104)   & 0.92 (0.89, 0.95) & 0.90 (0.87, 0.92) & 0.91 (0.88, 0.94) & 0.90 (0.87, 0.93) & 0.90 (0.86, 0.92)          \\
Enlarged Cardiom. (n=253) & 0.69 (0.65, 0.74) & 0.30 (0.26, 0.35) & 0.73 (0.68, 0.77) & 0.55 (0.51, 0.60) & \textbf{0.85 (0.81, 0.88)} \\
Lung Opacity (n=264)       & 0.90 (0.88, 0.93) & 0.92 (0.89, 0.94) & 0.89 (0.87, 0.92) & 0.89 (0.87, 0.92) & 0.91 (0.89, 0.94) \\
Support Devices (n=261)    & 0.89 (0.86, 0.92) & 0.89 (0.86, 0.92) & 0.85 (0.82, 0.88) & 0.86 (0.82, 0.89) & 0.87 (0.84, 0.90)          \\
No Finding (n=62)          & 0.89 (0.86, 0.93) & 0.88 (0.84, 0.91) & 0.89 (0.86, 0.92) & 0.89 (0.86, 0.92) & 0.92 (0.89, 0.95)          \\ \hline
Average                    & 0.84 (0.82, 0.85) & 0.79 (0.78, 0.80) & 0.84 (0.82, 0.85) & 0.81 (0.80, 0.83) & \textbf{0.88 (0.86, 0.89)} \\
Weighted Average           & 0.83 (0.81, 0.84) & 0.76 (0.74, 0.77) & 0.83 (0.81, 0.84) & 0.79 (0.77, 0.81) & \textbf{0.87 (0.86, 0.89)}
\end{tabular}%
}
\end{table}